# OPTIMIZATION OF ELECTROSPINNING TECHNIQUES FOR THE REALIZATION OF NANOFIBER PLASTIC LASERS

L. Persano*[a], M. Moffa[a], V. Fasano[b], M. Montinaro[b], G. Morello[a], V. Resta[a,b], D. Spadaro[c], P. G. Gucciardi[c], O. M. Maragò[c], A. Camposeo[a,d], D. Pisignano[a,b]

[a]Istituto Nanoscienze-CNR, Euromediterranean Center for Nanomaterial Modelling and Technology (ECMT), via Arnesano, I-73100 Lecce (Italy); [b]Dipartimento di Matematica e Fisica "Ennio De Giorgi", Università del Salento, via Arnesano, I-73100 Lecce (Italy); [c]CNR-IPCF, Istituto per i Processi Chimico-Fisici, viale F. Stagno D'Alcontres 37, I-98158 Messina (Italy); [d]Soft Materials and Technologies SRL, via Arnesano 16, I-73100 Lecce (Italy).

## ABSTRACT

Electrospinning technologies for the realization of active polymeric nanomaterials can be easily up-scaled, opening perspectives to industrial exploitation, and due to their versatility they can be employed to finely tailor the size, morphology and macroscopic assembly of fibers as well as their functional properties. Light-emitting or other active polymer nanofibers, made of conjugated polymers or of blends embedding chromophores or other functional dopants, are suitable for various applications in advanced photonics and sensing technologies. In particular, their almost one-dimensional geometry and finely tunable composition make them interesting materials for developing novel lasing devices. However, electrospinning techniques rely on a large variety of parameters and possible experimental geometries, and they need to be carefully optimized in order to obtain suitable topographical and photonic properties in the resulting nanostructures. Targeted features include smooth and uniform fiber surface, dimensional control, as well as filament alignment, enhanced light emission, and stimulated emission. We here present various optimization strategies for electrospinning methods which have been implemented and developed by us for the realization of lasing architectures based on polymer nanofibers. The geometry of the resulting nanowires leads to peculiar light-scattering from spun filaments, and to controllable lasing characteristics.

## 1. INTRODUCTION

In recent years, electrospinning technologies have emerged as reliable methods for realizing polymer-based nanomaterials, raising an increasing interest [1-4]. Through these approaches, fibers with diameter well below 1 μm can be aligned in arrays [5,6], deposited onto practically any substrate, or produced in the form of free-standing mats and with good throughput, which is appealing in view of industrial applications. Sectors of interest in this respect are, among others, sensors and bio-sensors in which polymer nanofibers can constitute active elements, actuators, energy harvesting by nanogenerators embedding piezo-fibers, smart clothing, and so on. What is more interesting for nanoelectronics and nanophotonics is the chemical versatility of electrospinning, which is largely compatible with conjugated polymers and with other active materials and light-emitting dyes which can be used as dopants. Electrospun light-emitting nanofibers include species made by polyphenylene vinylene (PPV) derivatives [7], polyfluorenes [8], polythiophenes [9], and by a large variety of transparent matrices such as poly(methyl methacrylate) (PMMA) and polystyrene (PS) filled by low-molar-mass dyes. However, many active compounds show a poor viscoelastic behavior or unsatisfying solubility, making a careful choice of the solution concentration and process optimization critical for obtaining nanofibers. More in general, electrospinning processes are apparently simple, but they depend on many parameters involving the physical and chemical properties of polymer solutions (used solvents, concentration, rheology, etc.), the applied bias and flow rates, as well as the processing environment (particularly humidity [10]). While these aspects make some polymer species hard to spin, they can become actually challenging for conjugated and light-emitting materials. In addition, though in electrospinning polymers are processed entirely at room temperature, the realized nanomaterials might be very sensitive to oxidation pathways, whose activation is favored by the high surface-to-volume ratios of the spun polymer filaments. For these reasons, electrospinning needs to be carefully optimized in order to obtain suitable topographical and photonic properties in the resulting nanostructures. Properties to achieve in turn depend on the specific functionality to be addressed. In particular, for realizing plastic lasers by electrospun luminescent fibers, one may be interested in controlling and possibly minimizing surface roughness, carefully tailoring the fiber size and arrangement, which in turn impact on the waveguiding capability of emitted photons supporting stimulated emission, and optimizing the emission yield. Due to these complex requirements, electrospun nanofibers exhibiting amplified spontaneous emission (ASE) and lasing are still rarely investigated [11-13]. Segments of individual nanofibers have been demonstrated, which emit in the visible with lasing excitation threshold fluence as low as 60 μJ cm$^{-2}$ [11]. Non-cylindrical dye-doped PS fibers have shown features of three dimensional etalon-like modes [12]. Randomly-distributed cavities, namely ring resonators, have been found in electrospun networks [13]. Distributed feedback geometries have been also explored, texturing the surface of individual electrospun fibers by periodic features obtained by nanoimprint lithography [14]. It is very likely that, by virtue of their chemical versatility and good degree of control in fiber assembling, electrospinning can be used as fabrication method helping in designing and realizing new and exotic lasing structures in the broad field of complex photonics.

*luana.persano@nano.cnr.it; phone +39 0832298118; fax +39 0832298146; www.nanojets.eu

Here, we focus on strategies based on electrospinning methods which have been implemented and developed in our group, aiming to realize new light-emitting materials and lasing architectures based on polymer nanofibers. For instance, processing conjugated polymers in nitrogen might lead to light-emitting nanofibers with higher photoluminescence quantum yield, and with better waveguiding performances [15]. Embedding dyes in plastic matrices such as PMMA is another, effective route to provide chromophores with a mechanically stable and protective coating, which are advantageous to limit photo-oxidation and to partially confine light along the fiber longitudinal axis, thus favoring stimulated emission. In this way, line-narrowing assisted by ASE can be achieved throughout the near-ultraviolet, visible, and near-infrared spectral range. Furthermore, the geometry of the resulting nanowires leads to peculiar light-scattering features from the spun filaments, which are here analyzed for individual fibers. The resulting light-light (*L-L*) characteristics of nanofiber-based plastic lasers with Rhodamine 590 Chloride show excitation threshold fluence of a few mJ cm$^{-2}$, depending on the specific peak within the found, multi-mode lasing spectra.

## 2. METHODS

Electrospinning of conjugated polymers in nitrogen is carried out by means of a glove box (Jacomex, GP[Concept]) equipped with a galvanic cell oxygen sensor and a $H_2O$ ceramic sensor, and keeping atmosphere $O_2$ below 2 ppm and $H_2O$ below 5 ppm, respectively. In another experiment, the solution for electrospinning is prepared dissolving 200 mg mL$^{-1}$ of poly(vinylpyrrolidone) (PVP, molecular weight = 1300 kDa, Alfa Aesar) in ethanol (Carlo Erba) and adding 1% w:w (dye:polymer) of Rhodamine 590 Chloride (molecular weight = 465 Da, Exciton). All the steps of solution preparation, stirring for 12 h at room temperature, and electrospinning are carried out in nitrogen atmosphere. Both the solution concentration and the dye/PVP relative content are chosen in order to achieve uniform fibers with good net gain. The solution is transferred in a syringe and held at a constant flow rate of 1 mL h$^{-1}$ by a peristaltic pump (33 Dual Syringe Pump, Harvard Apparatus Inc.). A bias of 11 kV is applied at the metal needle (21 gauge) of the syringe, employing a high-voltage power supply (EL60R0.6–22, Glassman High Voltage). Electrospun nanofibers are collected on a grounded metal collector and on 1 cm$^2$ quartz substrates.

Rhodamine 590 Chloride/PVP electrospun fibers are optically pumped by the 2$^{nd}$ harmonic of a pulsed Nd:YAG laser (Quanta-Ray INDI, Spectra-Physics), with 5-8 ns pulse duration and 10 Hz repetition rate. The excitation region on the samples is in the form of a stripe with length of about 3 mm and width of 70 μm, obtained by a cylindrical lens (focal length: 100 mm). Samples are kept in air during measurements, and photoluminescence is filtered with a 532 nm longpass filter (Razor Edge, Semrock with 2.7 nm transition width) and collected along the stripe direction by an optical fiber bundle with 0.22 numerical aperture. The emission is then dispersed via a monochromator (iHR320, Horiba) with a spectral resolution of 0.1 nm, coupled to a Peltier-cooled charge coupled device (Symphony, Horiba) with a 1024×256 pixel array. The measurements are performed with a 1200 grooves mm$^{-1}$ diffraction grating, using a 100 ms integration time. A beam splitter, put along the laser beam path, allows for real-time monitoring of the excitation energy through a pyroelectric detector (J3-09, Coherent). Analogous set-up configurations are used to collect ASE from polymer nanofibers doped by other dyes, with emission peaks ranging from the near-ultraviolet to the near-infrared. Scanning electron microscopy (SEM) is carried out by using a FEI Nova NanoSEM 450 system, with 10 kV acceleration voltage.

Confocal microscopy is carried out using a laser-scanning microscope relying on an inverted microscope (Eclipse Ti, Nikon) and a spectral scan head (Nikon).

## 3. RESULTS AND DISCUSSION

Figure 1 shows electrospun nanofibers produced in a nitrogen atmosphere and made of a yellow-emitting conjugated polymer (Fig. 1a) and of a red-emitting conjugated polymer blended with PVP (Fig. 1b). Average diameters of the fibers range from about 400 nm to about 2 μm, depending on the polymer. Following SEM inspection, these nanofibers appear generally smoother than samples of analogous composition and realized in air, which might lead to more efficient waveguiding of self-emitted light along the longitudinal axis of the polymer filaments [15]. We believe that a reduced incorporation of oxygen during the electrospinning step may effectively contribute in suppressing subsequent photo-oxidation paths. Confocal micrographs of these fibers highlight generally bright emission, which is quite uniform along the wire length, and the possible formation of beads (Fig. 1c,d). Analogous results are obtained for PMMA nanofibers doped with different dyes, as displayed in Figure 2a-c, although these nanowires generally exhibit more regular morphologies. Polymer nanofibers channel a larger fraction of emitted light along their length, due to the higher refractive index compared to surrounding air or substrate material underneath, a mechanism which can assist ASE. Indeed, ASE is generally observed in lasing dye-doped PMMA fibers (Fig. 2d). Regardless of the specific sample, emission spectra are broad, with full width at half maximum (FWHM) up to 90 nm at low excitation fluences, which is indicative of spontaneous emission, whereas the signal increases correspondingly to a significant line-narrowing at high excitation fluences (FWHM < 20 nm). Thresholds for line-narrowing are of the order of 0.1 mJ/cm$^2$ up to a few mJ/cm$^2$, depending on the nanofiber species. Spectral narrowing is known to be strictly related to ASE, in turn driven by photons travelling across a region of optical gain [16]. The emitted light intensity, $I_L$, is then given by:

$$I_L = \frac{I_p A(\lambda)}{G(\lambda)} \cdot \left[ e^{G(\lambda) \cdot L} - 1 \right], \qquad (1)$$

where $I_p$ is the pump intensity, $A(\lambda)$ is related to the spontaneous emission cross-section, $L$ is the length of the excited region and $G(\lambda)$ is the net gain. In this respect, optimizing the relative content of the dopants within the polymer is crucial for achieving optical gain, since stimulated emission can be strongly disfavored by clustering of active molecules and by too high dye concentrations. To overcome this issue, preliminary and extended experimental campaigns in which the dopant concentrations are systematically varied around values of 1-2% in weight are generally needed, which allows each species of light-emitting fibers to be ultimately realized with optimal composition. Modeling tools have also been made available, which allow the behavior of electrified jets and the size of the resulting nanofibers to be estimated once the spinning parameters are set [17]. Due to the tight interplay with sample thickness, the prediction of waveguiding-assisted ASE effects can be made easier by such estimations. These computational tools have been mainly validated through comparison with experiments using polymer solutions exhibiting a good viscoelastic behavior such as those based on PVP. However they can also offer valuable information for electrospinning processes involving active molecules given that the doping dyes are used in low enough concentration, thus not significantly affecting the rheology of the polymer solution.

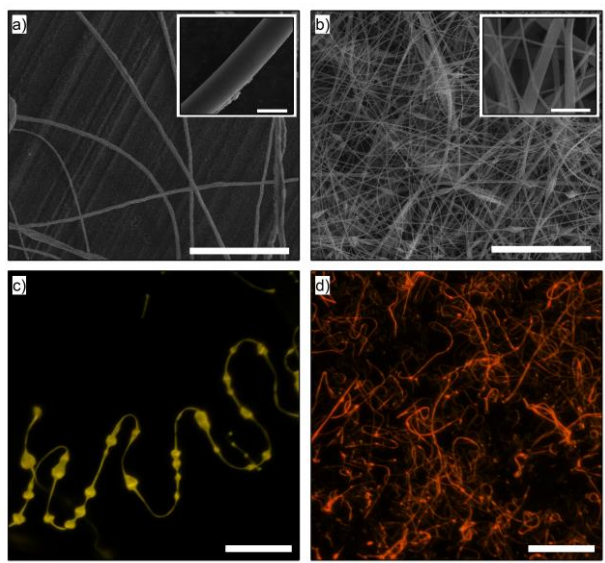

Figure 1. (a, b) Exemplary SEM micrographs of fibers made of a yellow-emitting polymer (a) and of a blend comprising a red-emitting polymer (b), electrospun in a nitrogen atmosphere. Scale bars: 50 μm (a) and 10 μm (b), respectively. Insets: Close-up of fibers. Scale bars: 5 μm (inset of a) and 1 μm (inset of b), respectively. (c, d) Corresponding confocal fluorescence micrographs. Scale bars: 100 μm. Excitation wavelengths: 408 nm (c) and 488 nm (d).

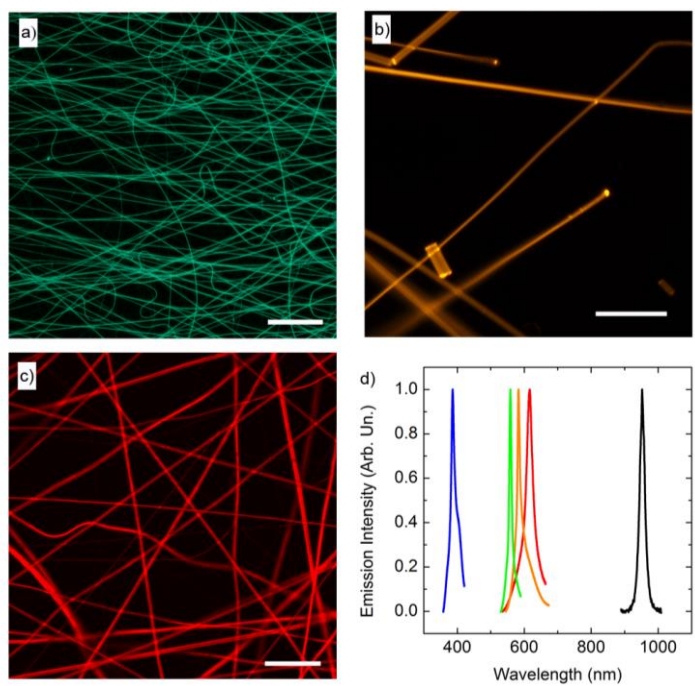

Figure 2. (a-c) Confocal fluorescence micrographs of nanofibers embedding different light-emitting molecules: (a) a pyrromethene complex, (b) [2-[2-[4-(dimethylamino)phenyl]ethenyl]-6-methyl-4H-pyran-4-ylidene]-propanedinitrile (DCM), (c) Rhodamine 590 Chloride. Scale bars: 50 μm. Excitation wavelengths: 408 nm (a,b) and 488 nm (c). (d) ASE spectra from nanofiber species embedding different dyes. The spectra are normalized to their maximum intensity. From lower to higher wavelengths, embedded molecules are a near-UV-emitting dye, a pyrromethene complex, Rhodamine 590 Chloride, DCM, and an infrared-emitting dye, respectively.

Together with waveguiding, light-scattering mechanisms play a major role in determining photon outcoupling as well as in affecting the eventual formation of disordered cavities in samples made by multiple filaments. Understanding how a light beam interacts with a nanostructured sample made by many filaments with size comparable or smaller than the wavelength of impinging light is challenging from both the experimental and the theoretical point of view, made difficult by the complex structure of the nanofibrous material. However, some important effects can be outlined by the behavior of light interacting with individual nanofibers. Without loss of generality, some assumptions can be done considering the specific features of polymer nanofibers in terms of size and optical properties, which allow their scattering properties to be investigated and some relevant relationships between the fiber geometrical parameters and the properties of diffused light to be unveiled. Here we model individual polymer nanofibers as infinitely long, dielectric cylindrical structures. The behavior of diffused light is rationalized under the Rayleigh-Gans approximation. This model applies when the following conditions are met:

$$|n-1| \ll 1 \quad (2a)$$

$$ka|n-1| \ll 1 \quad (2b)$$

where $n$ is the refractive index of the fiber relative to the surrounding medium, $k=2\pi/\lambda$ is the light wavevector, and $a$ is the characteristic size of the scattering particle, which in the case of a cylindrical object is its transverse radius, $a$. The Rayleigh-Gans model is obtained when the field inside a particle can be approximated by the incident field and hence the particle is not a strong scatterer. For instance, for polymer fibers with typical refractive index, $n \approx 1.5$, emission wavelengths in the visible range and size of the order of few hundreds of nanometers, the conditions set by Eq.s (2) are reliably met. Under these approximations one can obtain analytical expressions for the light scattering form factor, $f(\theta)$, for a given angle of incidence [18]. The intensity of the scattered light is proportional to $f(\theta)$, and, consequently, determining the light-scattering form factor allows useful information to be obtained about the angular distribution of the intensity of light diffused by a nanofiber. Moreover, by considering the incident light normal to the nanofiber longitudinal axis, for scattering directions in the plane normal to the fiber length the form factor takes a particularly simple expression:

$$f(\theta) = \frac{2J_1[2ka\sin(\theta/2)]}{2ka\sin(\theta/2)}, \qquad (3)$$

where $J_1$ is the Bessel function of the first kind with $l=1$ [19]. Figure 3 shows how the light scattering form factor varies for different values of the incident wavelength, $\lambda$, for a fixed nanofiber radius $a=100$ nm (Fig. 3a), and in terms of the polymer nanofiber radius, $a$, at a fixed wavelength $\lambda=633$ nm (Fig. 3b). $f$ is plotted as a function of the angle, $\theta$, in a plane perpendicular to the fiber longitudinal axis, as schematized in the inset of Fig. 3a. Longer wavelengths are more efficiently diffused at large angles (>90°) (Fig. 3a) and an increase of the light diffused at large angles is also achieved by decreasing the fiber size (Fig. 3b).

These aspects are relevant in order to design new materials with tailored optical and scattering properties. In particular, they are important to maximize the optical coupling between adjacent fibers by multi-scattering, an effect that can be exploited in many applications, including random lasers [20]. An exemplary room-temperature emission spectrum of Rhodamine 590 Chloride/PVP electrospun fibers under ns-pulsed excitation at a wavelength of 532 nm, is displayed in the inset of Figure 4. While at low excitation fluence (below about 6 mJ/cm$^2$) the emission spectra are generally featureless, for higher pump fluences various narrow peaks with FWHM below 2.5 nm) arise. In the inset of Fig. 4, at least two peaks can be appreciated at wavelengths, $\lambda_{L1}=594.7$ nm and $\lambda_{L2}=600.9$ nm, which are indicated by a square and by a circle, respectively, in the Figure. These peaks dominate over residual and lower modes. The corresponding *L-L* plots, in which the emitted intensity at the wavelengths corresponding to the two modes is plotted as a function of the incident excitation fluence, are shown in Fig. 4. The input-output characteristics evidence a roughly linear dependence of the emitted intensity on the excitation fluence above threshold, and different slopes for the two modes. These findings support the occurrence of complex light-scattering effects in the nanofibrous material, which can assist the onset of multi-mode lasing.

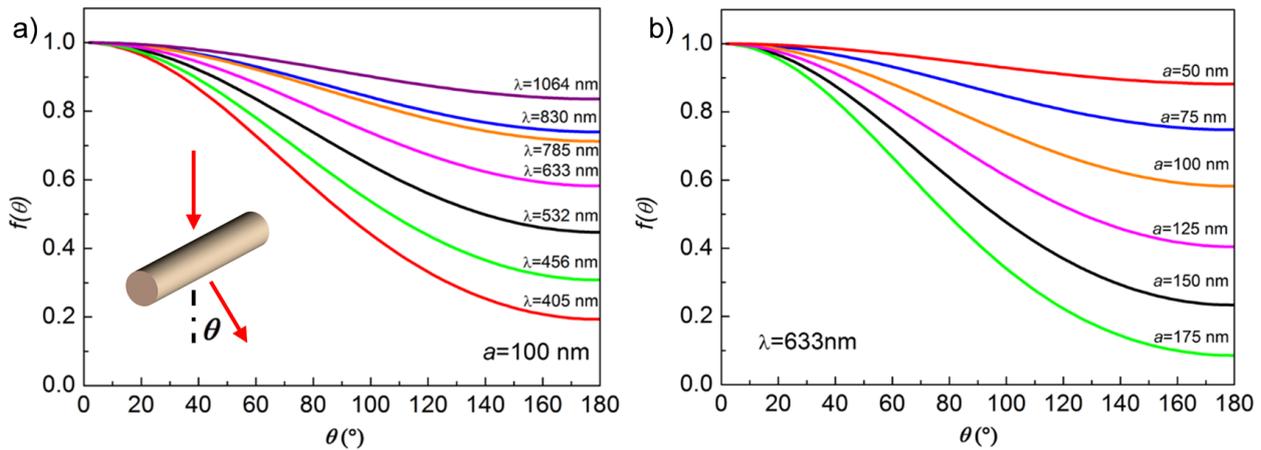

Figure 3. Calculated angular dependence of the normalized scattering form factor, $f(\theta)$, for different wavelengths of the incident light (a) and for different sizes of the nanofiber (b). The inset shows the configuration used for the calculation, namely light incident normally to the fiber longitudinal axis and light diffused in a plane perpendicular to this axis. Furthermore, the nanofiber is schematized as an infinite dielectric cylinder with radius $a$. The curves shown in (a) are obtained for a fiber having a radius $a=100$ nm, whereas in (b) a wavelength $\lambda=633$ nm is used for the incident light.

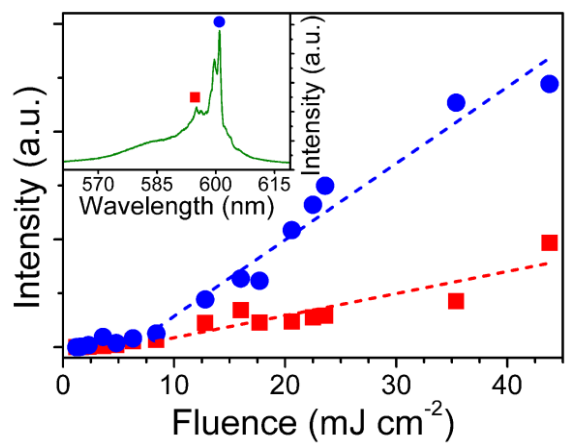

Figure 4. *L-L* plot of a nanofiber-based, multi-filament, multi-mode laser. The two highlighted modes exhibit a threshold pumping fluence of about 6 mJ/cm$^2$. Inset: single-shot spectrum measured at an excitation fluence of 24 mJ/cm$^2$. Symbols indicate the two modes at $\lambda_{L1}$=594.7 nm (square) and $\lambda_{L2}$=600.9 nm (circle), respectively.

## 4. CONCLUSIONS

Lasing architectures based on multi-filament polymer nanofibers realized by electrospinning show a significant potential for integration with microfluidic chips, sensors, and microsystems. In addition, the fabrication throughput reached by these technologies for generating light-emitting nanofibers is good enough to enable applications in diverse sectors, once the methods are optimized to achieve stable luminescence and optical gain in the produced nanostructures. Various routes are being explored in this respect. Investigating emission properties of the resulting nanofibrous materials together with light-scattering features may lead to designing and realizing novel plastic lasers. Next steps along this direction will include the systematic study of these properties in multi-filament organic materials.

*Acknowledgments*. The research leading to these results has received funding from the European Research Council under the European Union's Seventh Framework Programme (FP/2007-2013)/ERC Grant Agreement n. 306357 (ERC Starting Grant "NANO-JETS"). The authors also acknowledge the Apulia Regional Projects 'Networks of Public Research Laboratories', WAFITECH (09) and M. I. T. T. (13), and the Italian Ministry of Education, University and Research (MIUR), "Programma Operativo Nazionale Ricerca e Competitività" 2007-2013 e "Piano di Azione e Coesione", project PAC02L3_00087 "SOCIAL-NANO".